\documentclass[3p,times,procedia]{elsarticle}
\usepackage{nupha_ecrc}

\volume{00}

\firstpage{1}

\journalname{Nuclear Physics A}

\runauth{}

\jid{nupha}

\jnltitlelogo{Nuclear Physics A}

\usepackage{amssymb}
\usepackage[figuresright]{rotating}
\usepackage{color}
\usepackage{graphicx}
\usepackage{amsmath}
\newcommand{\diff}{\mathop{}\!\mathrm{d}}
\usepackage{siunitx}
\begin{document}

\begin{frontmatter}



%% Instructions from Editor: Please use the following \dochead only in the preprint version (e-print arXiv etc.); 
%% use empty \dochead{} when submitting to Nuclear Physics A!
\dochead{XXVIth International Conference on Ultrarelativistic Nucleus-Nucleus Collisions\\ (Quark Matter 2017)} %Address this
%% e.g. \dochead{17th International Conference on Dynamical Processes in Excited States of Solids}

\title{Holographic Jet Shapes and their Evolution in Strongly Coupled Plasma}


\author[label1]{Jasmine Brewer \footnote{Speaker}}
\author[label1]{Krishna Rajagopal}
\author[label1]{Andrey Sadofyev}
\author[label1,label2]{Wilke van der Schee}

\address[label1]{Center for Theoretical Physics, MIT, Cambridge MA 02139, USA}
\address[label2]{Institute for Theoretical Physics and Center for Extreme Matter and Emergent Phenomena, Utrecht University, Leuvenlaan 4, 3584 CE Utrecht, the Netherlands}

\begin{abstract}
Recently our group analyzed how the probability distribution for the jet opening angle is modified in an ensemble of jets that has propagated through an expanding cooling droplet of plasma \cite{Rajagopal:2016uip}. Each jet in the ensemble is represented holographically by a string in the dual 4+1- dimensional gravitational theory with the distribution of initial energies and opening angles in the ensemble given by perturbative QCD. In \cite{Rajagopal:2016uip}, the full string dynamics were approximated by assuming that the string moves at the speed of light. We are now able to analyze the full string dynamics for a range of possible initial conditions, giving us access to the dynamics of holographic jets just after their creation. 
The nullification timescale and the features of the string when it has nullified are all results of the string evolution. This emboldens us to analyze the full jet shape modification, rather than just the opening angle modification of each jet in the ensemble as in \cite{Rajagopal:2016uip}. We find the result that the jet shape scales with the opening angle at any particular energy. We construct an ensemble of dijets with energies and energy asymmetry distributions taken from events in proton-proton collisions, opening angle distribution as in \cite{Rajagopal:2016uip}, and jet shape taken from proton-proton collisions and scaled according to our result. We study how these observables are modified after we send the ensemble of dijets through the strongly-coupled plasma.
\end{abstract}


\end{frontmatter}

% \linenumbers

\section{Introduction}
\label{intro}

The discovery that the quark-gluon plasma created in heavy-ion collisions at RHIC and the LHC is strongly coupled has generated immense theoretical interest and leaves many unanswered questions. Jets in heavy ion collisions can provide important insights into QCD and the quark-gluon plasma, since they provide access to the interactions of hard partons with the medium and incorporate physics at widely separated momentum scales. Holography has emerged in recent years as an important tool for studying the strongly-coupled quark gluon plasma. Although QCD does not have a dual theory in holography, and the available theory $\mathcal{N}=4$ SYM is not asymptotically free, holography is nonetheless an important place to look for qualitative insights into how jets may interact with a strongly coupled plasma.
In this work, we construct an ensemble of holographic jets with initial energy and opening angle distributions taken from perturbative QCD, and study how that ensemble is modified after it propagates through an expanding, cooling droplet of strongly coupled plasma. We calculate the jet shape and dijet asymmetry modifications of our ensemble by the plasma and compare our results to data measured by CMS in Refs~\cite{Chatrchyan:2013kwa} and \cite{Chatrchyan:2012nia}.

\section{The Model}
\label{sec:model}

We consider a back-to-back pair of light quark jets losing energy as they propagate through an expanding, cooling droplet of strongly coupled $\mathcal{N}=4$ SYM plasma. In the gravitational description, the dynamics of a pair of light quark jets is described by an open fundamental string whose endpoints shoot away from each other and then fall into a black hole in an additional dimension in Anti-de Sitter (AdS) spacetime. The depth of the black hole in the AdS direction sets the temperature of the plasma in the field theory, and string energy which falls into the black hole in the gravitational description is energy lost by the quark-antiquark pair to thermalization by the plasma. 
The 5-dimensional metric in AdS which corresponds to a constant-temperature plasma in the 4-dimensional $\mathcal{N}=4$ SYM theory on its boundary is
\begin{equation}
	\label{metric}
	\diff s^2 = \frac{L^2}{u^2} \left(-f(u) \diff t^2 + \diff\vec{x}_\perp^{\:2} + \diff z^2 + \frac{\diff u^2}{f(u)}\right)\,,
\end{equation}
where $u$ is the additional direction in AdS space, $f(u)=1-u^4/u_H^4$, and the black hole is located at $u_H=1/\pi T$. Here $\vec{x}_\perp$ and $z$ are field theory coordinates specifying the transverse plane and the beam direction, respectively. This metric is an exact solution to Einstein's equations for a constant-temperature plasma, but for a spatially-varying temperature profile this model neglects transverse flow, fluid viscosity, and gradients.

\begin{figure}[!tbp]
  \centering
  \begin{minipage}[c]{0.48\textwidth}
    \includegraphics[width=\textwidth]{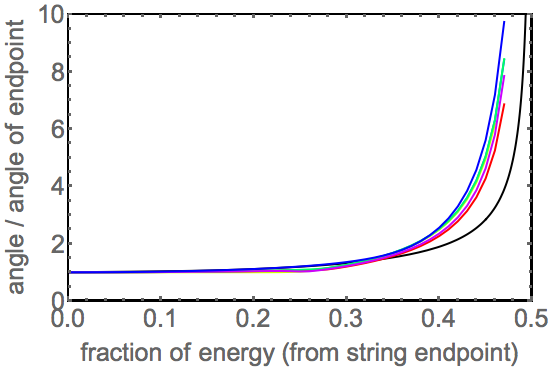}
    \caption{Distribution of energy $e(\sigma)$, parameterized by the opening angle $\sigma_0$, for two initial velocity profiles along the string prior to nullification and values of $E$ and $u_0$ spanning two orders of magnitude. The estimate for $e(\sigma)$ which was computed in a near-endpoint expansion in Ref~\cite{Chesler:2015nqz} and used in Ref~\cite{Rajagopal:2016uip} is shown in black for comparison.}
	\label{fig:univcurve}
  \end{minipage}
  \hfill
  \begin{minipage}[c]{0.48\textwidth}
    \includegraphics[width=\textwidth]{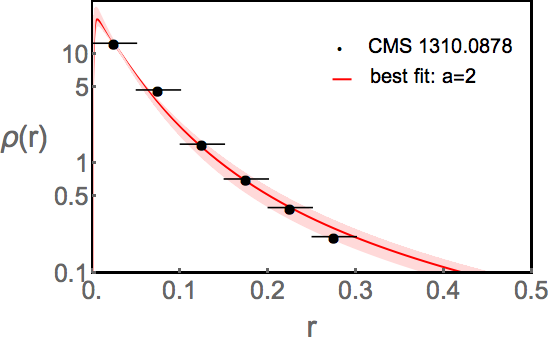}
    \caption{Jet shape in vacuum computed from the properties of nullified strings in holography and fit to CMS data on the jet shape in proton--proton collisions from Ref~\cite{Chatrchyan:2013kwa} shown as black symbols. The pink band shows a range in the free parameter $a=1.8-2.5$, with $a=2$ shown in dark red.}
	\label{fig:rhoplot}
  \end{minipage}
\end{figure}

\emph{Nullification of strings in vacuum.---} We are interested in the distribution of energy along the string while it is in the plasma, since this is what governs energy loss. Strings are known to become null as they fall, and after this the distribution of energy along them stays fixed for all time. With this as motivation, we first consider the initial dynamics and study the equilibration of real strings in vacuum. The dynamics of real classical strings are governed by the classical Nambu-Goto action, which specifies the motion given initial conditions on the position and velocity of the string. We tried several profiles for the initial velocity distributions, starting at fixed holographic depth $u_0$ with varying values of the initial angle of the endpoint $\sigma_0$ and varying values of the amplitude $E$ spanning two orders of magnitude. After these strings nullify we find that, as shown in Figure~\ref{fig:univcurve}, for all these families of initial conditions the shape of the energy distribution as a function of the AdS angle does not depend strongly on our choice of initial conditions as long as we rescale both energy and angle, so that they are measured with respect to the total energy and the opening angle of the endpoint.

\emph{An ensemble of jets in plasma.---} An individual jet in holography always widens when it propagates through plasma, since the angle of the endpoint is a proxy for the width of the jet in the field theory and the endpoint curves toward the black hole in the gravitational description. It was found in \cite{Rajagopal:2016uip}, however, that the competing effect that wider jets lose more energy may cause an \emph{ensemble} of jets to have an average opening angle which either narrows or widens. This suggests that considering an ensemble of jets is important for qualitative predictions of jet phenomenology in holography.
A useful measure of the opening angle of a jet is the variable 

\begin{equation}
\label{eqn:c11}
C_{1}^{(1)} = \sum_{i,j} z_i z_j \frac{\theta_{ij}}{R}\,,
\end{equation}
where the sum is over all pairs of hadrons in the jet, $\theta_{ij}$ is the angle between hadrons $i$ and $j$, and $z_i$ is the momentum fraction of hadron $i$. We take the jet radius parameter to be $R=0.3$ for consistency with CMS data. The distribution of $C_1^{(1)}$ has been calculated in perturbative QCD in \cite{Larkoski:2014wba}. The opening angle of a holographic jet is given by the angle of the string endpoint, $\sigma_0$, to leading order, but we do not have a direct analog of $C_1^{(1)}$ since the holographic calculation does not have hadrons and we cannot calculate Eq.~\eqref{eqn:c11} explicitly. Therefore we take $C_1^{(1)}= a \sigma_0$ for $a$ a free parameter in our model and fix $a$ by comparing to measurements of the jet shape in proton--proton collisions. We also take a distribution of initial jet energies which falls as $E^{-6}$. We take a simple blast-wave profile to model the temperature evolution in the transverse plane and assume boost-invariant longitudinal expansion (See Ref.~\cite{Rajagopal:2016uip} for details). We parameterize the differences in degrees of freedom and couplings between $\mathcal{N}=4$ SYM and QCD by rescaling their temperatures $T_{\mathcal{N}=4} = b T_{QCD}$, with $b$ the second free parameter in our model. We take the initial position of the quark-antiquark pair in the transverse plane to be distributed according to a binary scaling distribution proportional to the participant density, with randomly distributed direction.

\emph{Constraining the model from proton--proton data.---} With the distribution of jet energies and opening angles in perturbative QCD as described and a parametrization of the distribution of energy along the string in terms of the opening angle, we are able to compute the jet shape in vacuum. We utilize the result of \cite{Chesler:2014jva} for the angular distribution of power on the boundary of a null string falling in the AdS geometry:

\begin{equation}
\label{power_out}
\frac{dP_{out}}{d \cos r}=\frac{1}{2} \int d\sigma \frac{e(\sigma)}{\gamma(\sigma)^2 [1-v(\sigma) \cos r]^3}\,.
\end{equation}
Here $e(\sigma)$ is the worldsheet energy density $E =\int d\sigma e(\sigma) $, $\gamma(\sigma)=(1-v(\sigma)^2)^{-1/2}$, and $v(\sigma)=\cos \sigma$ for a null geodesic. The domain of integration is over the angles $\sigma$ along the worldsheet for which the null geodesics do not fall into the black hole. For a given value of the angle of the endpoint $\sigma_0$, we take $e(\sigma)$ to be given by the distribution of energy along real strings after nullification. The differential jet shape for an individual jet is the power $P_{out}$ as a function of the angle $r$ from the jet axis, as given by the integral of Eq.~\eqref{power_out}. The total differential jet shape is the average of the individual jet shapes over the ensemble. We fix the free parameter $a$ in our model by comparing this to CMS data on the jet shape in proton--proton collisions. The result shown in Figure~\ref{fig:rhoplot} has the best fit $a=2$ shown in red, which is in reasonable agreement with the crude estimate of $a \sim 1.7$ given in \cite{Rajagopal:2016uip} for smooth jets. We fix $b=0.203$, which gives a reasonable value of $R_{AA}^{jet} \approx 0.4$ for the number of jets of a given energy which emerge from the plasma over the number of jets of that energy which went in.

\section{Results and Conclusions}

\begin{figure}[!tbp]
  \centering
  \begin{minipage}[c]{0.48\textwidth}
    \includegraphics[width=\textwidth]{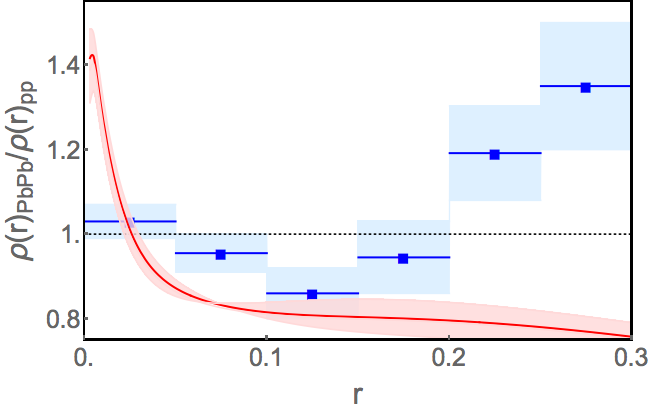}
    \caption{The jet shape modification of our ensemble of jets after propagation through the strongly-coupled plasma of Section \ref{sec:model} is shown in red for $a=2$, with the pink band indicating $a=1.8-2.5$. The CMS measurement of the jet shape modification in Pb--Pb collisions from \cite{Chatrchyan:2013kwa} is shown in blue symbols. The discrepancy between our calculation and data at large $r$ is expected, since we do not include the particles coming from the wake of the jet in the plasma that are then reconstructed as part of the jet \cite{Casalderrey-Solana:2016jvj}.}
	\label{fig:modrhoplot}
  \end{minipage}
  \hfill
  \begin{minipage}[c]{0.48\textwidth}
    \includegraphics[width=\textwidth]{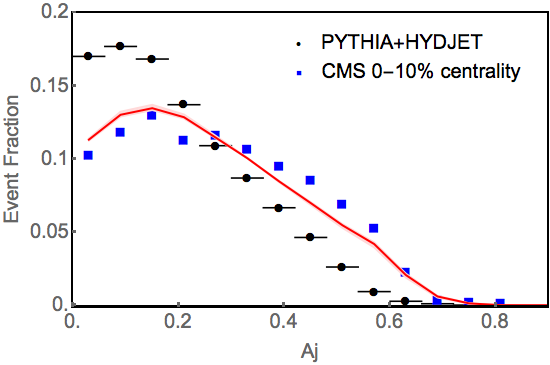}
    \caption{The dijet asymmetry modification of our ensemble of holographic jets after propagation through the strongly-coupled plasma of Section \ref{sec:model} is shown in red for $a=2$, with the pink band indicating $a=1.8-2.5$. We took the PYTHIA+HYDJET simulation from \cite{Chatrchyan:2012nia}, shown in black, as the initial distribution of dijet asymmetry for our ensemble. The CMS measurement of the dijet asymmetry modification for $\SIrange{0}{10}{\percent}$ centrality Pb--Pb collisions from \cite{Chatrchyan:2012nia} is shown in blue symbols.}
	\label{fig:dijetmodplot}
  \end{minipage}
\end{figure}

When the ensemble of jets with jet shape shown in Figure~\ref{fig:rhoplot} propagates through the expanding, cooling droplet of strongly-coupled plasma described in Section~\ref{sec:model}, individual jets widen and lose energy. We start the hydrodynamic profile at $\tau = \SI{1}{fm/c}$ after the collision and assume that the strings are null at that time and give the vacuum jet shape as matched to CMS data in Section~\ref{sec:model}. For simplicity, we assume that the string starts from a point at this time. The null strings propagate in plasma until the local temperature has dropped below the freezeout temperature $T_{FO}=\SI{175}{\MeV}$, after which they once again propagate in vacuum and their angles and energies no longer change. We calculate the jet shape modification between proton--proton and heavy ion collisions, $\rho(r)_{PbPb}/\rho(r)_{pp}$, from the modification by the plasma compared to the vacuum jet shape. For consistency with the CMS data on the jet shape modification given in Ref~\cite{Chatrchyan:2013kwa}, we impose the cut $p_T^{jet} > \SI{100}{\GeV/c}$. We compare the results of our calculation to the CMS data of \cite{Chatrchyan:2013kwa} in Figure~\ref{fig:modrhoplot}, finding qualitative agreement at small $r$. At larger $r$ our model does not include the 
soft particles coming from the 
wake in the plasma -- which carry the momentum lost by the jet and contribute to the reconstructed jet~\cite{Casalderrey-Solana:2016jvj}.

Another signature of the effect of the plasma on jets in heavy ion collisions is the significant enhancement of the asymmetry in dijet events. The dijet asymmetry is defined as $A_J=(p_{T,1}-p_{T,2})/(p_{T,1}+p_{T,2})$, where $p_{T,1}$ and $p_{T,2}$ are the transverse momenta of the leading and subleading jets, respectively. Due to the interplay between the presence of multiple jets, the substructure of jets, and the central role of jet-finding algorithms in the analysis of jets in both proton-proton and heavy ion collisions, the dijet asymmetry is not zero and the jets in a dijet are not back-to-back even in proton-proton collisions. This aspect is not naturally captured in our model of dijet events since we consider perfectly back-to-back strings. Instead, we construct an ensemble of 
%dijets whose asymmetry and acoplanarity distributions are 
back-to-back dijets whose asymmetry distribution is
as measured in proton-proton collisions by CMS in \cite{Chatrchyan:2012nia}, using a half of one of our strings to represent each jet in each dijet pair, and calculate the modification of this asymmetry distribution by propagation through plasma. We take $p_{T,1}>\SI{120}{\GeV/c}$ and $p_{T,2}>\SI{30}{\GeV/c}$ for consistency with Ref \cite{Chatrchyan:2012nia}. The results are shown in Figure~\ref{fig:dijetmodplot}, compared with CMS data from the $\SIrange{0}{10}{\percent}$ centrality bin. 
We find qualitative agreement between the dijet asymmetry modifications computed in this simple holographic model and data taken by CMS in heavy ion collisions. We anticipate that the largest systematic effect not represented in the pink band in Figure~\ref{fig:dijetmodplot} arises from the absence of three-jet events in our calculation, since these are in fact the origin of much of the dijet asymmetry in proton-proton collisions.  We plan to incorporate holographic three-jet events into our ensemble in future work.

\emph{Acknowledgements.---} This work was supported by the U.S. Department of Energy under grant Contract Number DE-SC0011090. (MIT-CTP/4903)

% \section{Conclusions}



%% References


\bibliographystyle{elsarticle-num}
% \begingroup
% \raggedright
\bibliography{nupha}
% \endgroup





% \end{thebibliography}

\end{document}